\documentclass[thmsa,a4paper,onecolumn,12pt,final,oneside,notitlepage]{article}
\usepackage{amssymb}
\usepackage{amsmath}
\usepackage{sw20lart}

\input{tcilatex}
\begin{document}

\title{Objective information in the empiricist view of von Weizs\"{a}cker}
\author{Iman Khatam \and Afshin Shafiee\thanks{%
Corresponding authur: shafiee@sharif.edu} \\
%EndAName
{\small Research Group On Foundations of Quantum Theory and Information, }\\
{\small Department of Chemistry, Sharif University of Technology,}\\
{\small P.O.Box 11365-9516, Tehran, Iran.}}
\maketitle

\begin{abstract}
We analyze von Weizs\"{a}cker's view regarding the concept of information in
physics. In his view, information arises from the reduction of properties of
a physical object to their logical descriptive propositions. The smallest
element of a lattice of propositions is an atom of information which is
considered as the essence of every physical identity including position
space. von Weizs\"{a}cker calls this element, \textquotedblleft
ur\textquotedblright . Moreover, Biological evolution is described in terms
of enhancement of the variety of forms. Form could be also reduced to
descriptive logical propositions, thus to atoms of information. Therefore,
information is the fundamental basis in von Weizs\"{a}cker's plan for
unifying all branches of Physics including Chemistry and Biology. Yet, there
are inadequacies in his lines of reasoning, critically assessed in this
paper.
\end{abstract}

\section{Introduction}

Carl Friedrich von Weizs\"{a}cker was a German philosopher and physicist
whose main research activities were on nuclear fusion in the sun and other
stars and on the creation of solar system. Yet, an impressive part of his
works which is the subject of the present paper, includes his efforts for
getting a unified description of all laws of physics under a theory called
ur-theory. He introduced ur-theory in \textquotedblleft Die Einheit der
Natur\textquotedblright\ in 1971[1]. An improved version of the book
entitled \textquotedblleft the structure of physics\textquotedblright\ has
been published in 1985 which was translated to English in 2006 [2]. This
book is the fruit of years of efforts for obtaining a theory of unity of
physics. In this book, information is the most fundamental concept
underlying physics. The centrality of information in physics, however, was
originally introduced by works on Boltzmann's formulation of entropy in
thermodynamics that was very similar to Shannon's measure of information and
Maxwell's demon which resulted to great works on the role of information in
physics. In the quantum area the concept of information was introduced by
von Neumann, before the recent work of von Weizs\"{a}cker. Weizs\"{a}cker's
innovation is in using information as a central concept for unifying all
branches of physics including Biology and chemistry which are also
considered to be reducible to physics and in defining \textquotedblleft
ur\textquotedblright s as atoms of information. Thus, information is
believed to be also a basic concept in the mentioned domains of natural
science. The first part of the book is devoted to the plan of reconstruction
of quantum theory which is the essence of ur-theory. In the second part of
the book, the role of information in unifying Biological evolution with the
Second Law of Thermodynamics is explained. In the last part of the book, von
Weizs\"{a}cker talks about interpretations of quantum theory.

In this paper we present a survey on his views focusing on the concept of
information. Since his philosophy has played a significant role on his view
about natural sciences and his plan for reconstruction of physics, we begin
with a short introduction about von Weizs\"{a}cker's philosophical attitude,
before entering and scrutinizing the main subject of his ideas, i.e.,
ur-theory. Concepts like experience, time and phenomenon have significant
roles in his terminology. Therefore, we have devoted the second part of our
paper to these concepts. In the third section, we explain his views about
reconstruction of quantum theory which is associated with the development of
ur-theory and its implications. In this regard, the role and significance of
the concept of information become clear. In section 4, we investigate the
same concept in Thermodynamics and Biology, in addition to von Weizs\"{a}%
cker's view on the consistency of Biological evolution with the Second Law
of Thermodynamics. Here, we assess the strength and coherence of his
insight. In section 5, we present a critical analysis of ur-theory and show
that the reduction of physical concepts to informational properties in the
way von Weizs\"{a}cker describes has fundamental problems. The conclusion
part is presented in the last section.

\section{Experience, phenomenon and time}

von Weizs\"{a}cker starts the first chapter of his book, \textquotedblleft
the structure of physics\textquotedblright\ by the basic statement
\textquotedblleft physics is based on experience\textquotedblright . This
statement has a central role in the conceptual structure of his theory. The
first point in his theory is gaining knowledge about the realm of
appearances of things or their phenomenological level. He is influenced by
Kant in this view point. The other point is the presumption of the concept
of time as a primary and fundamental concept which is a prerequisite of
experience in Kant's view as well. Kant assumed time and space as two 
\textit{a priori} features of experience, while they are not in themselves
perceivable in experience. Indeed, time and space are two subjective
concepts and every physical object corresponds to these subjective
intuitions by the process of gaining knowledge about natural phenomena,
where mind has a primary role here. Yet, we expect that each subject does
conform to a real independently existed object. On the other hand, Kant
believed that mind recognizes the thing not in itself, but through its
appearances, thus, through perception [3].

In the same way, von Weizs\"{a}cker thinks that our knowledge of physics is
possible based on the realm of appearances of physical entities
(observables) and by means of perception. We construct our knowledge about
an object using descriptions about their observables. Thus, necessarily we
presume the existence of an object with all of its observables to construct
knowledge about it. The object is matter and at the same time it has a form.
However, von Weizs\"{a}cker goes a step further respecting Kant who is
leaving aside space and excluding time as a precondition of experience. For,
von Weizs\"{a}cker states that space is derivable from ur-theory and it is
not a necessary precondition of experience.

According to von Weizs\"{a}cker, experience means learning from the past for
the sake of future. The past, present and future tenses are thus
prerequisites for experience [2, p.3]. Thus, experience is a temporal
concept. Similar to Kant who introduced intuition and subjective
prerequisites for empirical knowledge, which are known as Kant's twelve
categories, von Weizs\"{a}cker assumes time order as a necessary condition
for empirical knowledge.

As mentioned earlier in Kant's view, the object conforms to subject by
gaining knowledge about natural phenomena, instead of that the subject
conforms to the real independently existed objects. von Weizs\"{a}cker is
apparently influenced by this viewpoint in constructing his ur-theory.

When one talks about an event in the past, thenceforth, (s)he considers it
as a factual event. However, events pertaining to future have not actually
occurred yet and are accounted as potentialities. In this regard,
probability by its very nature is a temporal concept. This seems to mean
that, one can ascribe a probabilistic statement only for events pertaining
to the future, while, we cannot talk about factually past occurred events
with probabilistic propositions. Therefore, the concept of probability is
intertwined with the concept of upcoming experiences. Hence, the concepts of
time and probability are considered as prerequisites of empirical natural
science \footnote[1]{%
We will not use the adjective "empirical"\ before "science" throughout our
work in this paper, for in von Weizs\"{a}cker's view natural science is
generally empirical}. Regarding the importance of temporal experience in
Weizsacker's view and the distinction between past and future in temporal
events, one should notice that, however, the meaning of \textquotedblleft
factually past occurred events\textquotedblright\ is completely vague in
quantum mechanics. For example, let us assume that at $t=t_{0}$, we measure
the spin component of a spin-%
%TCIMACRO{\U{bd} }%
%BeginExpansion
$\frac12$
%EndExpansion
particle alonge $z$-direction, $S_{z}$, and obtain $+\frac{\hbar }{2}$.
Then, we perform a similar measurement along $x$-direction, $S_{x}$, at $%
t=t_{1}>t_{0}$. What can we say about $S_{z}$ at $t_{1}$? It is a past
occurred event with value $+\frac{\hbar }{2}$ , but it is not factual in any
sense at $t_{1}$. Any fact is a fact in quantum events, when we know it with
certainty. Otherwise, it is an indeterminate phenomenon. So, our experience
is limited to perceived facts at present, and past events are only important
for making predictions about the future status of the system. At present,
nothing can be said about the actual values obtained before for the system.
Hence, our experiences at different times do not form a shared experience as
a whole. For classical objects, also, there are situations in which one
cannot talk about a past occurred event with certainty, because it might be
measured with low precision. Past events, even if considered factual could
be probabilistic.

The aforementioned lattice of propositions is established about properties
which could be known by experience or by the realm of appearances of a given
object. This is the basis for one of the main postulates of von Weizs\"{a}%
cker in establishing ur-theory, i.e., the existence of separable
alternatives to which the next section is devoted.

\section{Ur-theory}

Ur-theory was a result of von Weizs\"{a}cker's efforts towards establishing
a unified theory of physics. Because physics is the most fundamental of all
natural sciences including chemistry and Biology, such a theory should be
capable of describing these domains of science as well. In his viewpoint,
quantum theory among all theories of physics is the most fundamental and
comprehensive one which could be the base for such a unified physical
theory. The probabilistic nature of predictions of quantum theory, also,
makes it general. One could assume quantum theory as a general theory of
probability for its statistical predictions and von Neumann's abstract
mathematical description of the theory can be viewed as a mathematical basis
for a fundamental theory of physics [2, p.9].

von Weizs\"{a}cker considers quantum theory as possessing two levels. One is
an abstract level which is limited to the mathematical framework of von
Neumann, in which everything is described in Hilbert space without
considering actual and classical concepts like position space, potentials,
fields and particles. These concepts are pertaining to the second level,
which von Weizs\"{a}cker calls the concrete level of the quantum theory.
Then, he reconstructs an abstract representation of quantum theory using
some postulates to establish his ur-theory [2, 4-9].

We previously said that in von Weizs\"{a}cker's view, quantum theory is a
general theory of probability. This theory helps us to analyze everything in
terms of binary alternatives with yes/no answers which mean that there exist
empirically decidable alternatives. This is the first and most significant
postulate of ur-theory. These alternatives could be either possibilities of
the occurrence of an event or propositions with which one describes a
phenomenon which could be empirically tested. Indeed, what von Weizs\"{a}%
cker actually does is reducing physical observables to logical concepts of
alternatives. When we want to gain empirical knowledge about something, we
assume a number of logical alternatives about that object or event. Any
repeatable experimental result is one of these alternatives. In this case,
we say that observable properties are reduced to logical alternatives.
However, this is possible only if one presupposes the existence of the
object. This postulate is applicable in classical mechanics as well. One
could divide every concept to smaller and smaller descriptive elements by
means of binary alternatives until reaching the smallest indivisible part
which is called by von Weizs\"{a}cker \textquotedblleft ur\textquotedblright
. Actually, this postulate follows the classical logic in which such an
analysis is applicable for every perceivable property of an object. The only
difference between classical and quantum mechanical logic in this regard is
the selection of the corresponding lattice of propositions which is related
to the definition and significance of information in each domain. One could
deduce that for indivisible phenomena, two inseparable systems should be
considered as a united whole. In other words, in these cases we are dealing
with a non-classical phenomenon with a quantum mechanical nature. Here the
word \textquotedblleft phenomenon\textquotedblright\ is brought from Bohr as
a result of his efforts for resolving the measurement problem and the effect
of the measurement device on observed results. Bohr thought that the system
and the measurement setup should be treated as one inseparable whole, known
as phenomenon, considering that the underlying nature of interaction between
the measuring setup and the system is in principle unknowable. Following
this line of thought, the lattice of propositions in classical mechanics is
established about separable objects and in quantum mechanics, it is
established about phenomena. Thus, the only difference between classical and
quantum mechanical probabilities as mentioned earlier is in the selection of
the lattice of propositions.

Therefore, the lattice of propositions is not exclusively applicable to
separable events, but also to inseparable events. For instance, the
entanglement between spatially separated quantum particles, results in a
correlation between them in such a way that the information of one particle
becomes dependent on the information of the other particle. This leads to an
excess of information in quantum mechanics, relative to classical mechanics,
because entanglement provides exact knowledge of one of the systems without
any measurement performed on the other system. Similarly, one can mention
the quantum uncertainty relation in which one cannot obtain sharp
distributions for two incompatible observables in a simultaneous
measurement. This means that complete and exact information about both
incompatible observables is not accessible, simultaneously. In the plan of
reconstruction of quantum theory, these points should be taken seriously.
Taking these points into consideration, von Weizs\"{a}cker introduces his
second postulate, namely the postulate of indeterminacy or postulate of
expansion [2, p.77]:

\begin{quote}
To any two mutually exclusive final propositions $a_{1}$ and $a_{2}$ about
an object, there is a final proposition about the same object which does not
exclude either of the two. Two propositions $x$ and $y$ exclude one another
if $p(x,y)=p(y,x)=0.$
\end{quote}

The quantum theoretical excess information is a result of this postulate.
This is one of the necessary requirements of any theory of probability, as
he denotes. In an expansion, no term possesses an absolute true value and as
we mentioned earlier, probability is assumed as a temporal concept here,
meaning that one can use probabilistic statements only for future events.
Thus, in his opinion, time is the only concept amongst all of classical
concepts that has a significant and fundamental role in the reconstruction
of abstract quantum mechanics and is assumed by von Weizsacker as a
realistic variable instead of being a parameter. In this regard, inspired
from classical fundamental concept of time, he introduces the third
postulate namely the postulate of \textquotedblleft
dynamics\textquotedblright . However, contrary to classical mechanics in the
development of ur-theory, there are real distinctions between future and
past modalities of time. This distinction reveals itself in the
irreversibility of the measurement process and in the reduction of the wave
function. Moreover, von Weizsacker tries to prove the Second Law of
Thermodynamics using Boltzmann's H-theorem and through his definition of
time. He believes that time also reveals itself as a real physical entity in
Thermodynamics. We leave this point for further consideration in future.

As is apparent, two abstract prerequisites for experience are identifiable
in von Weizs\"{a}cker's view. The first one is the concept of time that is
defined as the real distinction between the past and the potential future.
The second one is the existence of decidable alternatives. Each alternative
can be analyzed and divided into binary alternatives called
\textquotedblleft ur\textquotedblright s. This decidability is based on
experience. One could plan experiments in which one should choose an answer
between two opposite results, i.e., yes and no. Step by step, this analysis
reaches an ultimate proposition with yes/no decisions. This final
irreducible alternative is called \textquotedblleft ur\textquotedblright .
Based on these premises, von Weizs\"{a}cker establishes a lattice of logical
binary propositions about every observable property. This lattice has $2^{k}$
dimensions in Hilbert space. If one uses classical information, i.e.,
Shannon information, this space possesses $k$ bits of information. Yet,
according to the expansion postulate, in quantum mechanics one could expand
the given state space in terms of basis states defined for incompatible
observables. Consequently, one can assign an unlimited number of bits of
information to this $2^{k}$ dimensional lattice of propositions. The
ultimate binary alternative \textquotedblleft ur\textquotedblright\ is the
smallest part of the lattice and the smallest element of physical
information as well which possesses spinor characteristics, meaning that it
has a two dimensional Hilbert space and could be assumed as an atom of
information with a substantial role in the structure of physics. Indeed,
every physical property which is empirically knowable to us could be reduced
to these atoms of information. Definitely, von Weizs\"{a}cker asserts that
one can construct all objective and (non-abstract) concepts of quantum
mechanics based on the concept of information and ur-theory. For instance,
the three-dimensional position space could be reduced to the concept of
\textquotedblleft ur\textquotedblright\ and its spinor space, accordingly.
The reasoning for this assertion is that the proper symmetry group of urs is
spinor's symmetry group of SU(2). Taking advantage of the isomorphism of
this symmetry group with the symmetry group of rotation in three dimensional
space (i.e., SO(3)), von Weizsacker deduced the three dimensional position
space from two dimensional Hilbert space of \textquotedblleft
ur\textquotedblright s. Moreover, he was aiming to deduce all other quantum
mechanical concepts like field and particle using the substantial concept of
atoms of information [10-12].

\section{The role of information in Thermodynamics and Biological evolution}

Evolution in this survey refers to the emergence of new forms in Biological
systems during the history of earth. If entropy is a quantitative measure of
irreversibility, information is the quantitative measure of evolution as
well. Information refers to the amount of form, in other words the amount of
Gestalt. Since Biological evolution is associated with increasing of variety
of forms, the amount of information increases in this process. The amount of
form is a property of an object and is defined by von Weizs\"{a}cker as the
following [2, p.214]:

\begin{quote}
The more decisions can be made about an object, the more form one can
recognize in it, in a general not necessarily spatial meaning of the word.
\end{quote}

In other words, the amount of form is equivalent to the number of the
possibilities for different descriptions of a given object. For example, a
DNA molecule is described by various possibilities for the arrangement of
its nucleotides, a spin-%
%TCIMACRO{\U{bd} }%
%BeginExpansion
$\frac12$
%EndExpansion
particle is described by two possibilities for its spin components and $N$
number of harmonic oscillators are described by $2N$ possibilities for their
parity\footnote[1]{%
Parity ($\widehat{\Pi }$) is a unitary operator in quantum mechanics with
two eigenvalues $+1$ and $-1$, defining in turn, even and odd parities for
an arbitrary state-vector $\left\vert \psi _{i}\right\rangle $. In a
mathematical description, $\widehat{\Pi }\left\vert \psi _{i}\right\rangle
=\pm 1\left\vert \psi _{i}\right\rangle $. As is obvious parity refers to
the symmetry of the wave function under space coordinate inversion.}, etc. 

On the other hand, Thermodynamic entropy is equivalent to the potential
information. Therefore, information is a fundamental concept in both domains
of Thermodynamics and Biology.

Because of the mentioned common characteristic of irreversibility in
Thermodynamics and Biological evolution, these two domains of nature are
assumed as one phenomenon occurring in two different levels. Here, however,
one encounters the controversial problem of whether the Biological evolution
occurs in accordance with the Second Law of Thermodynamics. However, von
Weizs\"{a}cker's answer is yes, Biological evolution doesn't violate the
Second Law of Thermodynamics. He believes that the mentioned problem arises
from the fact that in text books of Thermodynamics, the increase of entropy
is described as an increase of disorder. On the other hand, Biological
evolution is associated with increasing of order. Accordingly, these two
domains seem to be in contrast with each other. Nevertheless, one may
consider order or disorder as esthetical concepts, but not physical ones. To
consider biological systems, one should have into account non-equilibrium
thermodynamics in which the rate of dissipation of energy and time rate of
entropy production is crucial.

Regardless of this point, von Weizs\"{a}cker believes that [2, p.215]:

\begin{quote}
It is usual to interpret entropy as a measure of disorder, and thereby
thermodynamic irreversibility as an increase in disorder. Evolution,
however, is understood as an increase in possible forms, and in that sense
as order. Under these premises evolution must be perceived as a process
proceeding against thermodynamic irreversibility. Here, exactly the opposite
thesis is to be presented: Under suitable circumstances, an increase in
entropy is identical to the growth of forms; evolution is a special case of
the irreversibility of events.\textquotedblright
\end{quote}

Let us first see how von Weizs\"{a}cker defines information in these
domains. Information has not an absolute definition. As he says: An
\textquotedblleft absolute\textquotedblright\ concept of information has no
meaning; information exists only \textquotedblleft under one
concept,\textquotedblright\ more accurately, \textquotedblleft relative on
two semantic levels\textquotedblright \lbrack 2, p.217]. We need at least
two semantic levels for which information could be defined relatively. For
instance, in Statistical Thermodynamics, information is quantified relative
to macro-state and micro-state levels. The amount of information of a
macro-state is related to the number of micro-states it contains. Therefore,
we cannot talk about information of a macro-state without considering its
corresponding micro-states. Determination of the micro and macro-states
depends on the physical system (or the object) under study. For instance,
DNA chain constitutes the macro-state and different arrangements of its
nucleotides make its micro-states. Assuming that entropy is equivalent to
information, von Weizs\"{a}cker explicitly states that entropy is a measure
of form. The more information is there about an object, the greater is the
number of forms (possible descriptions) of that object. Based on this
argument, von Weizs\"{a}cker claims that assigning the increase of entropy
to the increase of disorder is a linguistic mistake. As he believes [2,
p.222]:

\begin{quote}
The concept of entropy is so general and abstract that the specification of
a high \textit{a priori} probability for a state rich in form also amounts
to assigning a high entropy to it.
\end{quote}

Therefore, the concept of information in Thermodynamics and Biology is
somehow similar.

However, one may question what is the difference between that information
which is the measure of irreversibility and that which is the measure of
evolution? Their only difference is in the selection of the micro-state and
macro-state levels. Additionally, there is a correspondence between entropy
and the amount of Gestalt. So, any modeling of Thermodynamic irreversibility
according to The Second Law regarding the increase of disorder is a
linguistic mistake and should be revised.

In Thermodynamics, one only needs two corresponding levels of
micro-state-macro-state to gain information from a system. But, in Biology,
von Weizs\"{a}cker adds another level to the Thermodynamic semantic levels,
called the morphological level. One measures the information of a
macro-state with respect to the micro-states and entropy or information
increases with the increase of the number of micro-states. In a similar way,
in Biology information of a morphological state is evaluated with respect to
its lower levels of macro and micro-states. Thus, von Weizs\"{a}cker
concludes that the both above processes, i.e., Thermodynamic irreversibility
and Biological evolution include the concept of information as a central
notion, so that both of them are substantially related to information. Thus,
they are identical in nature, but occur in different levels of natural
processes.

von Weizs\"{a}cker proves this claim using two lines of verification. The
first is a model that we analyze first. According to this model, if the
first level (i.e., the micro-state) is illustrated as the atomic level and
the second level or the macro-state is modeled by the molecular level, then
the level of variety of forms in Biology -- which includes a spectrum of
Biological macro-molecules to a diversity of living systems is pertaining to
the third level called the morphological level. In Thermodynamics, the
quantification of information associated with a process needs only the first
two semantic levels. While, in Biological evolution one needs the third
level in addition to the first and second levels. For example, the bases in
a DNA molecule make the atomic level, then the molecular level is the
arrangement of bases in a part of the DNA molecule making, e.g., the gene
which determines the color of eyes in a human body which constitutes the
morphological level.

The clearest explanation is better to be quoted directly from von
Weizsacker's book [2, p.222]:

\begin{quote}
Restricting attention, as in the previous examples, to two semantic levels
and one thereby defined concept of information, there follows from the
structure of time only the Second Law: as time progresses, the actual
information of the state present at that time will decrease with
overwhelming probability, its potential information (entropy) increase. If
one wants to express the development of form by means of the concept of
information at all, one must introduce (at least) three semantic levels,
with three different measures of information then defined among them.
\end{quote}

Consequently, Thermodynamic irreversibility is a special case of the
Biological evolution, occurring at lower levels and both processes include
the notion of information within their foundation.

The other line of verification, Weizs\"{a}cker talks about, is based on the
Second Law of Thermodynamics. He uses the growth of a crystal as an
illustration. This process is a good example of the increase of order in a
system. Since it is a spontaneous process, von Weizs\"{a}cker deduces that
it occurs with accordance to the second law and is associated with an
increase of entropy. Thus, an increase of entropy is associated with an
increase of order in the crystallization process. He finally concludes that
information increases in this process, and the same occurs in Biological
evolution, i.e., the increase of potential forms.

In a crystallization process, there are infinite possibilities for the
generation of a final form which show the variety of arrangements in both
molecular (macro-state) level and atomic (micro-state) level. By increasing
these possibilities, the amount of information increases as well. This is
analogous to what occurs in Biological evolution.

However, neither of the above arguments is flawless. First, he models the
three levels of micro-state, macro-state and the level of variety of forms
by atomic, molecular and morphological levels, respectively. Indeed he
models the levels of energy distribution of particles which have a crucial
role in Statistical Thermodynamics, with the variety of molecular forms
produced by combination of atoms. It seems that the Thermodynamic level
could be described by characterizing the particle's level and its underlying
classical laws. At the same time, atoms with appropriate energies combine
with each other to form molecules which are the more organized forms
relative to the atomic level. The Thermodynamic level including the assembly
of atoms (or the molecular level) is assumed to be a more organized level
than the atomic level in von Weizs\"{a}cker's view. A higher level of
organization occurs at the Biological level. While, the formation of
molecules from atoms could be understood in terms of atomic levels of energy
and the formation of molecular orbitals according to quantum mechanics,
Thermodynamic process is not reducible to the dynamical-mechanical
description. It seems that he wants to build a connection between
micro-states (described by quantum mechanics) with the atomic level (which
models the micro-states) on one hand, and the macro-state (described by
Thermodynamics) with the molecular level (which models the macro-state) on
the other hand in such a way that one could reduce the Thermodynamic level
to the dynamical-mechanical level. One could in principle establish
molecular structures based on dynamical-mechanical rules. Yet, the
Thermodynamics level is not reducible to these rules even in principle.
Since, e.g., entropy is not a dynamical-mechanical notion. With a closer
look, for gaining information about a macro-state one needs to consider the
number of micro-states. So, the variety and multiplicity are characteristic
features of the micro-level. For example, the mean value of energy in
Thermodynamics is a measure of temperature and is being used to describe the
macro-state. In this case, the micro-states are the individual particle's
energy levels.

In effect, the order of levels in Biological evolution according to von Weizs%
\"{a}cker's model is as the following,

$Atomic\rightarrow Molecular\rightarrow Morphological$

So, as is expected, the higher levels should be the more general levels and
the lower levels should possess more variety and multiplicity. Every higher
level should be considered as a macro-state with respect to its lower level.
Therefore, the morphological level should be considered as a macro-state
with respect to the molecular level, in the same way that the molecular
level is a macro-state with respect to the atomic level (micro-state) and
the amount of information in each case is evaluated with respect to
attributes of these levels.

Yet, one should notice that the multiplicity of forms in Biological
evolution appears at the morphological level. Then, the direction of the
relation between these semantic levels changes. Because, the macro-state in
Biology (corresponding to the morphological level) has more variety and
multiplicity than its corresponding molecular level. This relationship is
somehow similar to the relationship between the atomic and molecular levels.
For, at the atomic level, the multiplicity of states (or, correspondingly,
the abstract forms of knowledge) is again considerable. This is important,
when the notion of information is taken into account. So, the true direction
of levels changes to the following form:

$Atomic\rightarrow Molecular\leftarrow Morphological$

The molecular level seems to have the less diversity with respect to both
atomic and morphological levels. Consequently, to define information, the
hierarchy of levels with respect to each other is not identical in
irreversible processes and Biological evolution. Hence, the model doesn't
fulfill the real situation.

The second line of reasoning for unifying entropy and evolution is also not
acceptable. Although, it is true that the process of formation of a crystal
occurs in accordance with the Second Law of Thermodynamics, it is actually
associated with an entropy decrease for the system and not increase as von
Weizs\"{a}cker assumes. Nevertheless, the decrease of entropy in this
process is dominated by decreasing enthalpy and the crystallization occurs
spontaneously.

Had we assumed that von Weizs\"{a}cker considered the whole entropy of the
universe, this reasoning could still not be true [2, p.221]. Since, in the
latter case, one cannot refer to the properties of the system and its states
directly, as is usually requested for evaluating information.

\section{A critical analysis of ur-theory}

The work of von Weizs\"{a}cker in the difficult area of philosophy of
science is an appreciable work, for it was a great challenge in unifying
physics. It attracted much attention and resulted in branches of insights on
the understanding of the concept of information in natural sciences up to
the present time.

We saw through the three previous sections that von Weizs\"{a}cker considers
information as a fundamental concept which is the basis for everything in
physics. The physical object could be constructed from atoms of information.
Furthermore, in the same line of reasoning the position space, forces and
every physical observable could be reduced to an information basis. In
addition, we saw that in Thermodynamics and Biological evolution,
information identifies the amount of potential forms (and possibilities).
The more the required number of propositions for describing an object, the
more is its information and consequently, the more is its amount of Gestalt
(form).

One may ask, however, what von Weizs\"{a}cker means by \textquotedblleft
form\textquotedblright ? Following Aristotelian philosophy, he considers
form as the essence of everything. Yet, form is a property of the object and
knowable to us. This property is not separable from the object. Form, like
any other property, should be analyzed by informational propositions with
yes/no answers. The more propositions are required to describe an object,
the more are its possible forms. Thus, form is the substance and information
is the measure of substance. The ultimate alternative is called
\textquotedblleft ur\textquotedblright\ that is an atom of information.
Here, the classical atomism is replaced by an abstract logical atomism. In
the classical atomism, the division of an object to smaller parts takes
place in the position space. Divisibility in terms of logical
\textquotedblleft ur\textquotedblright s is assumed by von Weizs\"{a}cker as
a radical atomism which is an atomism based on the concept of information.

But, what is information? In addition to being a measure of form,
information is what is understood [2, p.304]. This kind of definition mixes
the subjective and objective notions of information all together. For
example, the arrangement of nucleotides in the DNA molecular structure
includes objective information about the future behavior of a living being
independent of whether it could be observed or not. \textquotedblleft
Understanding\textquotedblright\ in this case is the development of genotype
to phenotype according to von Weizs\"{a}cker. This means that when a given
form of a living being is changed in an evolution period to cause a
transition from genotype to phenotype, a kind of understanding is also
developed during this transition. This \textquotedblleft
understanding\textquotedblright\ is an attribute of the organism, itself,
not dependent on perception of another conscious being.

This phenotype is an objectified semantic. On the other hand, we need at
least two semantic levels of macro-state and micro-state to evaluate
information. Thus, information has not an absolute definition, here. The
alternatives or potentialities are not absolute but are defined relative to
special semantic levels. The same situation holds for information. The
definition of propositions or \textquotedblleft ur\textquotedblright s is
constrained by the restrictions of measurement instruments and limitations
of human knowledge. Information in von Weizs\"{a}cker's view is neither
matter nor energy. But, it is another entity independent from both of them.

Semantic levels of information encounter us with concepts of mind and matter
in von Weizs\"{a}cker's philosophy, for which there is no fundamental
duality in this view. Mind should be understood as a natural entity like
matter, and human knowledge is a process occurring in Nature and belonging
to it. So, knowledge is not independent of the matter stuff of the world.
Information is both objective and subjective at the same time. The
information about an object is objectively there, whether we know about it
or not. However, it is knowable to us.

An object has a remarkable ontological aspect. Its observables are
accidental or extrinsic properties and have no independent existence. Their
existence depends on object's existence. They are knowable to us and we
analyze them by logical propositions. This logical aspect of knowledge which
is based on perception, ultimately leads to a \textquotedblleft
ur\textquotedblright , i.e., the ultimate alternative of information.
\textquotedblleft Ur\textquotedblright\ is a logical concept and yet is not
independently definable without assuming an object and its observables.
Considering what von Weizs\"{a}cker says about the dependence of the
decision of the lattice of propositions or the definition of
\textquotedblleft ur\textquotedblright s on the limitations of measurement
instruments, \textquotedblleft ur\textquotedblright\ is relatively an
accidental concept.

It is admissible to ask how such a concept could be considered fundamental
and substantial. Unfortunately, the definitions of an object and the atom of
information called \textquotedblleft ur\textquotedblright\ make a loop.One
might consider the object and the atom of information as a whole, just like
the system and the measuring instrument in quantum mechanics. Yet, there
could be no \textquotedblleft ur\textquotedblright\ when there is no object.
Any proposition about an object dependes \textit{per se} on the nature of
the object being considered. Then, its nature refers to proposition being
used. First, an object is presumed, then, its observable properties are
defined through which one could gain knowledge about the object. Therefore,
one analyzes these observables by binary alternatives which could be
questions with equi-probable yes/no answers until reaching a final
proposition. However, the final proposition (or a \textquotedblleft
ur\textquotedblright ) depends on our previous decisions of propositions and
is not distinguishable from them by itself. For example a proposition might
be about up and down spin components of a quantum particle or about its
massiveness. Such a proposition depends on the nature of the object being
considered. So, information about an object is object-dependent. In another
words, it is claimed that a \textquotedblleft ur\textquotedblright\ is
substantial. Yet, the nature of an object should be presumed, because
without the definition of an object and characterizing its properties, the
definition of a network of propositions which could ultimately lead to an
atom of information \textquotedblleft ur\textquotedblright\ is impossible.
So, an irreducible element of subjectivism (i.e., the role of our minds in
knowing and characterizing the matter stuff of the world including all
objects and their properties) is introduced.

Regarding Kant's view, the scientific knowledge refers to properties of
physical objects as conceptualized by us. Our conceptions and our mental
frameworks influence our scientific knowledge. For instance, we naturally
have a Euclidean imagery of space. Our minds bear space and time as
prerequisites of experience and our apprehension occur within this mental
framework. This could be generalized to any logical method of gaining
knowledge. However, limiting natural sciences to human subjectivism is a
limitation to the creative and innovative apprehension of physical phenomena.

For example, one can notice that imaging a non-Euclidean geometry for space
in Relativity Theory of Einstein could not be understood in terms of mental
presupposition of special structures of space-time. So, Relativity Theory is
a good example for how mind conforms to the external reality and that how
physics does not match our mental framework. However, von Weizs\"{a}cker,
trying to avoid the loop of subjective-objective information, declares that
the process of cognition and gaining empirical knowledge is a process
belonging to Nature itself. In his view form is the essence of every object
and it is also the essence of mind. Nevertheless, if we accept this argument
we expect that a physical explanation for the unity of mind and matter in
terms of \textquotedblleft ur\textquotedblright s could be possible. No such
underlying physics has been given yet. Furthermore, ur-theory does not
explain how matter originates from the abstract logical concept of
\textquotedblleft ur\textquotedblright . From a different view, in von Weizs%
\"{a}cker's opinion, experience is the cornerstone of physics. Experience is
defined by him as the usage of data obtained from observation and
measurements occurred in past for doing predictions about the future. Or as
he says, experience means learning from the past for the sake of the future.
However, in some situations, we cannot know all properties of an object by
experience (or through observation). In such instances how can one establish
a lattice of propositions reducible to a final proposition or a
\textquotedblleft ur\textquotedblright ? For example, consider the known
problem of wave-particle duality in quantum mechanics. The quantum object
has a particle-like behavior in some experimental conditions and behaves as
a wave in other situations. Our experimental setups determine what we could
observe for the object. Then our decisions influence the final
\textquotedblleft ur\textquotedblright\ proposition which can tell the
object is a particle or wave. Each of these entities in turn has a
completely different lattice of propositions. The proposition about the
identity of being wave or particle in the logical reduction approach toward
a \textquotedblleft ur\textquotedblright , is unavoidably dependent upon our
experimental conditions. So, according to ur-theory wave or particle should
be assumed as a property of the whole experimental setup of a quantum object
and not its nature by itself. But, what does ur-theory tell about an
individual quantum object or its own nature? How can \textquotedblleft
ur\textquotedblright\ which is expected to be a fundamental and substantial
entity depend on the external and empirical conditions of observation at the
same time? The atom of information is context-dependent, just like the
physical behavior of the object, itself. So, what is the fundamental
preference of a \textquotedblleft ur\textquotedblright ? This leads to the
conclusion that an atom of information has a dual character in its essence,
corresponding to the dual character of a quantum object in different
situations. Then, what is more fundamental, an object or its information? In
this way, we will never know what a quantum object is and what its physical
nature implies. The problem remains a problem and our knowledge of physics
transforms to the knowledge of description of observations as it happened in
the early quantum theory. Replacing objective atoms with \textquotedblleft
ur\textquotedblright s does not solve any problem to help us obtaining a
clearer picture of Nature and the behavior of its parts.

On the other hands, as is apparent from ur-theory, atoms of information are
quantitative measures of a given object. It is evident that measures of the
properties of a physical object could not be the essence of that object,
especially if these measures only evaluate the extrinsic properties. This
makes science be limited to merely perceptible levels of phenomena.

von Weizs\"{a}cker's view which confines natural science to human perception
and the realm of appearances of physical things (observable properties) is
problematic. As Kant implies, the impossibility of substantive and priory
knowledge independent of experience leads to the conclusion that the object
must conform to the subject to make knowledge possible. Influenced by this
view, von Weizs\"{a}cker introduces knowledge about an object by classifying
observations in a logical subjective framework of mind and constructs a
lattice of propositions or atoms of information. One should note that,
however, there are theories in physics that have predictions proved
empirically after they are predicted by the theory itself. For example, the
existence of positrons was theoretically predicted by Dirac in quantum field
theory. It was four years after this prediction that the particle was
detected empirically by Carl Anderson. These kinds of theoretical
predictions affirm that making correspondence between the predictions of
theory with experience is not the mere way of gaining knowledge. Yet, in
ur-theory, von Weizs\"{a}cker concentrates on the role of experience. The
lattice of propositions about an object is constructed using the data
obtained from experiments or former observations.

This way that confines our knowledge to perception and empirical incomes
cannot resolve explanatory problems of quantum phenomena. In order to
explain quantum weirdnesses (like wave-particle duality, nonlocal
correlations, etc.) which are unexplainable by classical theories, using
perception is not sufficient at all. Leaving aside material atomism and
replacing it with informational atomism which is not more than a measure for
manifesting forms of an object resolves no paradoxes. Moreover, it leads to
new vague problems like: How could material objects be constructed from
atoms of information? Or how could one explain subject-object association in
natural events by an information-based view?

For information to be so fundamental and objective as von Weizsacker aimed
to set it, we need a more abstract definition of this concept. Information
needs to arise not from human logical roots. It should be independent of
semantics and free of transmission aspects of communication. A step toward
such concept was taken by Gornitz through introducing the objective concept
of protyposis [13]:

\begin{quote}
Physics is more than an \textquotedblleft extension of
logics\textquotedblright , and, in physics information differs from
destination, or meaning, or knowledge. Meaning always has a subjective
aspect too, so meaning cannot be a basis for science and objectivity. If
quantum information is to become the basis for science it must be conceived
as absolute quantum information, free of meaning. It is denominated as
\textquotedblleft Protyposis\textquotedblright\ to avoid the connotation of
information and meaning. Protyposis enables a fundamentally new
understanding of matter which can be seen as \textquotedblleft
formed\textquotedblright , \textquotedblleft condensed\textquotedblright\ or
\textquotedblleft designed\textquotedblright\ abstract quantum information.
\end{quote}

This concept is believed to be a basis for derivation of Einsteinian
structure of the non-Euclidian space-time [14].

Gornitz also offers an answer to our previously asked question by the
concept of protyposis that \textquotedblleft how could material objects be
constructed from atoms of information?\textquotedblright \lbrack 15]:

\begin{quote}
If the protyposis should be connected with the established parts of physics,
then, beside other requirements, the construction of relativistic particles
from qubits has to be given. This means that irreducible representations of
the Poincare-group must be constructed.
\end{quote}

However, it is not clear how different aspects of reality, like material
objects, energy, space-time, etc., could be obtained based on a unique and
abstract concept of protyposis. How do these variety of concepts arise from
such an abstract concept, without reference to any distinguishing agent
necessary for making these distinctions meaningful? Could it predict new
concepts or is it just a descriptive concept which provides a different
language by which everything could be reduced to protyposis?

One of the other main problems of ur-theory is that when a lattice of
propositions is constructed for an object, it should include all properties
of that object. This means that all possible propositions that could be used
for describing an object should be included in the logical lattice of
propositions, so that the greater is the number of propositions, the greater
is the number of forms that object possesses. Considering continuous
variables, like $x$ and $p$, however, there is no complete set of
propositions for describing the quantum state. This means that there are
possibilities which could not be reduced to a final proposition in
principle. So, a final proposition is not possible for an object and we are
faced with an infinite number of elements of a lattice of information. Then,
one cannot extract an object from informational basis. Yet, in von Weizs\"{a}%
cker's view, the existence of a final proposition is a necessary condition
for an object to actually exist[2, p.77]:

\begin{quote}
If a certain object actually exists, then a final proposition about it is
always necessary.
\end{quote}

As von Weizs\"{a}cker claims, this statement is equivalent to saying that
[2, p.77]:

\begin{quote}
Every object has at any time as property a probability distribution of all
its properties.
\end{quote}

However, the problem is that the existence of a lattice of propositions is
not always associated with the existence of a final proposition. In cases
which a final proposition does not exist for a system, von Weizs\"{a}cker
says that such an object does not actually exist at all. This is in
contradiction with situations described by continuous variables.

And at last, let's see the role of information in Biology again. Biological
evolution is associated with the increase of variety of forms and
Biodiversity. If one asks why evolution takes place in this way, the answer
is usually that this process occurs, because it is the more probable
process. This answer is the same as the answer to the question of why
Thermodynamic irreversibility occurs as the more probable process of
increasing entropy. In static states, we can say the form of Biological
systems could be described using a lattice of binary propositions. The final
proposition is a \textquotedblleft ur\textquotedblright\ with two equally
likely answers. This shows that this lattice is basically symmetric. But a
dynamic process like evolution --which occurs, because it is more likely to
occur- could not be explained using the fully symmetric theory of
information. In completely symmetric states, information reaches its maximum
value and could not further increase. Therefore, diversity of forms does not
occur as well. One may logically analyze a static system to reach a final
proposition, but application of ur-theory to dynamic changes like evolution
is paradoxical. Regarding information theory as a time-independent
representation, ur-theory makes no distinction between states before and
after evolution. Ur-theory is symmetric in the selection among several
possibilities. It does not explain how such a selection among states which
the system evolves toward them occurs. On the other hands, even if we accept
that applying ur-theory to the dynamics of a quantum system has no problem,
we have no unique mathematical model for the dynamics for a system under
evolution which is different with a mechanical system. So, one have no
answer according to ur-theory for biological evolution.

On the other hands, we know that there are cases in Nature for which such a
symmetry is broken. For example two chiral molecules have a completely
similar lattice of logical alternatives. But each final proposition is the
opposite alternative of the other one. Yet, in nature, one of these cases
uniquely occurs. For example, all amino acids which constitute natural
proteins are right-handed and all tao neutrinos are left-handed. Almost all
spiral oysters are right-handed. This break of symmetry is not
understandable here, because left-handed and right-handed answers are
equally probable according to ur-theory.

For continuous variables, however, no complete set of propositions exist.
For instance, with a discrete variable like spin, we have in every direction
a complete set of propositions given by up and down alternatives. For
continuous variables like $x$ and $p$, the number of alternatives are not
finite.

This is somehow similar to saying that for a continuous variable we have
analog information, while, for a discrete variable, every proposition
contains a digital information. Analog information is innumerable and
infinite, so it prevents us from constructing a lattice of propositions with
definite limits [16].

\section{Conclusions}

Our analysis of ur-theory shows that replacing material atomism by logical
informational atomism leaves quantum paradoxes unresolved, and yet is not
successful in unifying Biology, Thermodynamics and Quantum Theory.
Furthermore, it leads to additional obscurities like how logical elements of
information could build material objects in a physical sense.
\textquotedblleft Ur\textquotedblright s as atoms of information are
dependent on human decisions and confined to instrumental limitations and
could not be assumed substantial, as ur-theory implies. Another problem is
about applying ur-theory to Biology, especially in resolving the problem of
homo-chirality which bears a kind of symmetry breaking in nature. Even if we
consider there is another universe which has opposite symmetries to ours,
this doesn't explain why this kind of symmetry breaking or chairlity exists
in our universe.

\bigskip 

\textbf{Acknowledgement} \textit{The authors would like to thank the two
anonymous referees for their valuable comments which led to improvment of
our paper.}

\end{document}